\documentclass[10pt]{article}
\usepackage{geometry}                
\usepackage{graphicx}
\usepackage{subfig}
\usepackage{amssymb}
\usepackage{amsmath}
\usepackage{hyperref}
\usepackage{epstopdf}
\usepackage{caption}
\usepackage[round]{natbib}
\newcommand{\vecn}{\boldsymbol{\mathbf{n}}}

\newcommand{\vect}{\boldsymbol{\mathbf{t}}}

\newcommand{\vecw}{\boldsymbol{\mathbf{w}}}
\newcommand{\vecx}{\boldsymbol{\mathbf{x}}}
\newcommand{\vecy}{\boldsymbol{\mathbf{y}}}

\newcommand{\matX}{\boldsymbol{\mathbf{X}}}
\newcommand{\matY}{\boldsymbol{\mathbf{Y}}}

\newcommand{\pv}{\boldsymbol{\phi}}

\usepackage[T1]{fontenc}
\usepackage[utf8]{inputenc}
\usepackage{authblk}

\title{Quantification of tumour evolution and heterogeneity via Bayesian epiallele detection}
\author[1]{James E. Barrett\thanks{Contact: regmjeb@ucl.ac.uk}}
\author[1]{Andrew Feber}
\author[1]{Javier Herrero}
\author[1]{Miljana Tanic}
\author[2,3]{Gareth Wilson}
\author[2,3,4,$\dagger$]{Charles Swanton}
\author[1]{Stephan Beck}
\affil[$\dagger$]{on behalf of the Lung TRACERx consortium}
\affil[1]{UCL Cancer Institute, University College London, U.K.}
\affil[2]{The Francis Crick Institute, London, U.K.}
\affil[3]{Cancer Research U.K. Lung Cancer Centre of Excellence, UCL Cancer Institute, U.K.}
\affil[4]{University College London Hospitals NHS Foundation Trust, U.K.}
\date{Feburary 20, 2017}

\begin{document}
\maketitle

\begin{abstract}

\textbf{Motivation:} Epigenetic heterogeneity within a tumour can play an important role in tumour evolution and the emergence of resistance to treatment. It is increasingly recognised that the study of DNA methylation (DNAm) patterns along the genome -- so-called `epialleles' -- offers greater insight into epigenetic dynamics than conventional analyses which examine DNAm marks individually.\\
\textbf{Results:} We have developed a Bayesian model to infer which epialleles are present in multiple regions of the same tumour. We apply our method to reduced representation bisulfite sequencing (RRBS) data from multiple regions of one lung cancer tumour and a matched normal sample. The model borrows information from all tumour regions to leverage greater statistical power. The total number of epialleles, the epiallele DNAm patterns, and a noise hyperparameter are all automatically inferred from the data. Uncertainty as to which epiallele an observed sequencing read originated from is explicitly incorporated by marginalising over the appropriate posterior densities. The degree to which tumour samples are contaminated with normal tissue can be estimated and corrected for. By tracing the distribution of epialleles throughout the tumour we can infer the phylogenetic history of the tumour, identify epialleles that differ between normal and cancer tissue, and define a measure of global epigenetic disorder.\\
\textbf{Availability:} R code is available at \url{github.com/james-e-barrett}.\\
\textbf{Contact:} \href{regmjeb@ucl.ac.uk}{regmjeb@ucl.ac.uk}
\end{abstract}

%
%
\section{Introduction}
\label{sec:intro}
%
%

Epigenetic variability allows greater phenotypic diversity and plasticity within a population of genetically similar cells. Epigenetic diversity within a tumour provides a mechanism for clonal evolution and the emergence of resistance to therapy \citep{mazor2016intratumoral}. Persistence of treatment-resistant subclonal populations may explain the failure of some therapies, and higher levels of heterogeneity have been associated with poorer clinical outcomes \citep{landau2014locally}. Analysing multiple tissue samples from different tumour regions facilitates quantification of tumour heterogeneity and phylogenetic analyses. \citet{pan2015epigenomic} showed that intra-tumour DNAm heterogeneity is predictive of time-to-relapse in diffuse B-cell lymphomas. \citet{brocks2014intratumor} have shown that both epigenetic and genetic alterations reflect the evolutioary history of prostate cancers. A recent study of Ewing sarcoma found substantial levels of epigenetic heterogeneity within tumours \citep{sheffield2017dna}.

Epigenetic modifications play an important role in the regulation of gene expression. One of the most common types is DNA methylation (DNAm) --- where a methyl group is added to cytosine. We will focus on DNAm in the canonical CpG context where cytosine (C) is followed by guanine (G). High levels of DNAm in promoter regions are associated with suppressed gene expression whereas increased methylation in gene body regions tends to have the opposite effect \citep{suzuki2008dna}.

Reduced representation bisulfite sequencing (RRBS) is a sequencing technique that measures DNAm \citep{gu2011preparation}. The experimental protocol consists of treating DNA with bisulphite which converts unmethylated cytosines into uracils. During the amplification process uracils are converted into thymines. After sequencing and comparison to a reference genome, unconverted CpGs are identified as unmethylated and vice versa. The RRBS technique does not sequence the entire genome, but rather regions of the genome that are enriched for CpGs. This naturally splits the genome into distinct loci which can be analysed separately.

Conventional analyses of DNAm have focused on the average DNAm level per CpG site. This is obtained by examining all of the sequencing reads which contain a given CpG and simply counting how many times it is methylated. This type of analysis, however, fails to take into account the full methylation pattern at a given locus which can be observed by looking at all contiguous CpGs along a sequencing read. If there are $d$ CpG sites on one read then there are $2^d$ possible methylation patterns, which are called \emph{epialleles} \citep{richards2006inherited}. Sequencing reads that cover the same $d$ CpG sites can be compared, and the frequency of distinct epialleles that are present can be calculated. Since each DNA fragment comes from a different cell (more precisely a different allele) this provides a snapshot of how many distinct cellular subpopulations are present within the sample. The additional information acquired from contiguous CpG sites on sequencing reads is not present using array-based platforms. It is becoming clear that leveraging this extra information offers potential insights into the epigenetic landscape that would otherwise be missed \citep{li2014dynamic, lin2015estimation, he2013dmeas}.

If multiple samples are taken from the same tumour then each sample can be analysed to see which epialleles are present, and in what proportion, at a given loci. By tracing the presence and absence of different epialleles across different regions of the tumour and matched normal tissue it is possible to reconstruct the evolutionary history of the tumour regions, and to probe for significant differences between normal and tumour tissue. Moreover, the diversity of epialleles within the tumour provides a measure of overall epigenetic heterogeneity.

The acquisition of tumour samples may result in a mixture of both tumour and normal tissue. By comparing the expression of epialleles within the tumour samples and matched normal tissue it is possible to estimate the sample purity --- the proportion of the sample which is tumour tissue. Furthermore, it is possible to decontaminate the tumour samples by effectively `subtracting' that component of the epiallele profile which can be attributed to the contaminating normal tissue. An analysis of differential epiallele expression and phylogenetics can be conducted after decontamination.

We present a Bayesian statistical model to infer which epialleles are present at a given locus. The model infers the epialleles that are present and which epiallele each observed sequencing read corresponds to. One hyperparameter controls the level of noise in the model (which represents errors due to bisulfite conversion, PCR amplification, and sequencing) and this is also inferred from the data. Finally, the total number of distinct epialleles is inferred. This final step is a model selection problem and we use the Akaike Information Criterion to avoid overfitting the model. The Bayesian approach allows the quantification of uncertainty regarding the model parameters. In particular, there may be some ambiguity as to which epiallele a certain observed read corresponds to (if some epialleles are very similar to each other for instance). This uncertainty is incorporated into the epiallele distribution by averaging over the appropriate model parameters with respect to the corresponding posterior density.

\subsection{Related work}

The additional information garnered from adjacent CpGs can be used to define a measure of variability or heterogeneity within a biological sample. The concept of `epipolymorphism', for instance, has been proposed by \citet{landan2012epigenetic}. \citet{xie2011genome} define a measure of `methylation entropy' based on the Shannon entropy and \citet{landau2014locally} developed the concept of `proportion of discordant reads'.

The term \emph{allele-specific methylation} has also been used to refer to epialleles. Statistical models have been developed by \citet{peng2012detection, fang2012genomic, wu2015nonparametric} to identify epialleles at a given locus and which epiallele each observed read originated from. These models can infer multiple epialleles but in applications only two epialleles have been assumed. An algorithm to estimate tumour purity and deconvolve the epigenomes of tumour and normal tissue uses a very similar statistical model \citep{zheng2014methylpurify}.

\citet{li2014dynamic} compare the epiallele distribution at two disease stages using a `composition entropy difference calculation'. They identify loci with substantial shifts in epiallele composition. They confine their analysis to epialleles defined by four CpG sites. \citet{lee2015new} used multinomial logistic regression to test for differences in the epiallele distribution between normal and cancer cells. They report performance that is very similar to the method of \citet{li2014dynamic}, but do not constrain their approach to four CpGs. In both of these approaches the epialleles are identified from the raw sequencing data, without any inference step to account for experimental noise.

\citet{lin2015estimation} develop a statistical model that explicitly takes into account measurement noise due to bisulfite conversion and sequencing errors. The model allows identification of `spurious' epialleles that are due to measurement error (spurious epialleles will tend to have low counts and be very similar to a dominant epiallele). Noise parameters are manually estimated from experimental data, and missing data are not facilitated by their model.

In summary, an adequate epiallele analysis of DNAm sequencing data should have the following features. It should answer the basic research question of whether there is a difference in the epiallele composition between two or more groups of samples --- and identify the loci at which there are significant differences. Ideally, some measures need to be taken to avoid spurious epiallele detection due to experimental noise. In addition, an analysis method will generally need to accommodate variable sequencing depth per loci, a variable number of contiguous CpGs per sequencing read, and missing data. Missing data can arise from partially overlapping reads or gaps in a read due to non-overlapping paired-end sequencing protocols.

In addition to the above features, our Bayesian approach automatically infers all model parameters (including the total number of epialleles) from the observed data. Ambiguity in model parameters is explicitly incorporated in our analysis by averaging over the appropriate Bayesian posterior density (descried in detail below). We have applied our method to data from multiple tumour regions and matched normal tissue. We have developed a protocol for estimating the tumour sample purity and consequently decontaminating the inferred epiallele profiles. Although we have focused on multi-region tumour sampling our method could be applied to a single sample also.

\begin{figure*}
\centering
\includegraphics[scale=0.68]{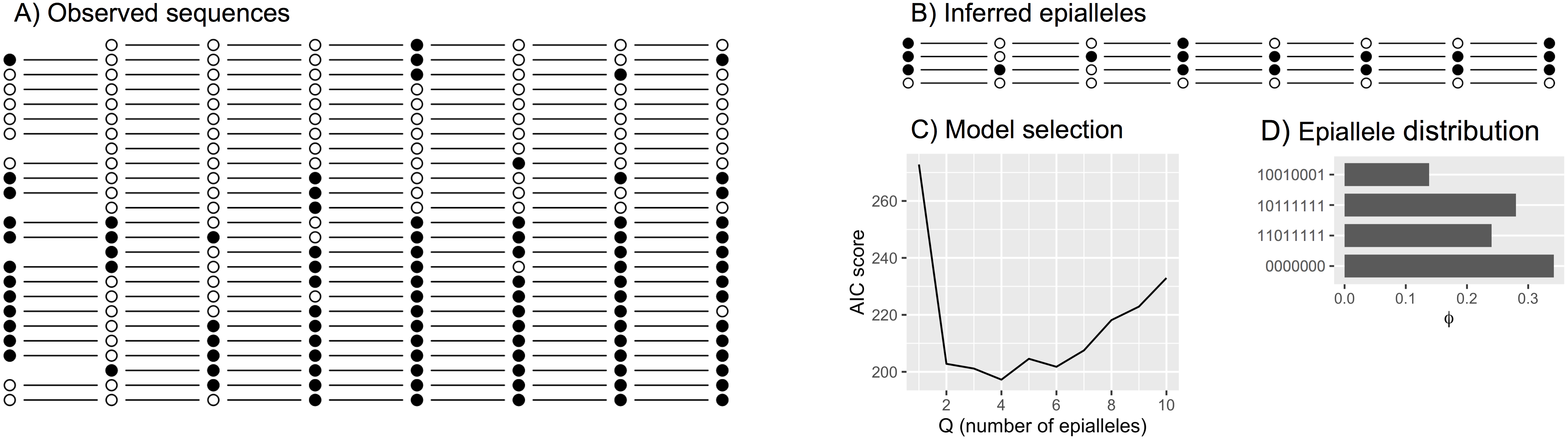}
\caption{(a) An example of a genomic locus (chr1:1,145,478-1,145,614) in which each row corresponds to a sequencing read. Black and white circles represent methylated and unmethylated CpGs respectively. Note that some CpG measurements are missing. (b) The four epialleles that are inferred from the observed sequencing reads. (c) The Akaike Information Criterion score versus the total number of epialleles. The inferred number of epialleles corresponds to the minimum AIC score. (d) The proportion of observed reads attributed to each epiallele after marginalisation over the parameter $\vecw$ (see main text for details).}
\label{fig:method}
\end{figure*}

%
%
\section{Methods}
\label{sec:methods}
%

Sequencing reads are aligned to the reference genome and organised into different genomic \emph{loci}. A locus is a region of the genome containing $d$ CpG sites ($d$ can take different values to each locus). Due to the nature of RRBS data the sequencing reads naturally tend to form non-overlapping loci. Some additional steps were taken to modify loci in order to control the amount of missing data per locus. See Supplementary Material A for full details.

Let $N$ denote the number of sequencing reads at a given locus. To keep our notation compact we will avoid indexing each locus and what follows here is applicable to any locus of the genome. A sequencing read is represented by a $d$-dimensional vector $\vecy_i\in\{0,1\}^d$ where $i=1,\ldots,N$ and 0 and 1 correspond to unmethylated and methylated CpG sites respectively. An example is plotted in Figure \ref{fig:method} (a). It is assumed that each observed read can be attributed to one of $Q$ epialleles $\vecx_q$ with $q=1,\ldots,Q$ and $Q\leq N$. The parameter $w_i\in(1,\ldots,Q)$ specifies which epiallele read $\vecy_i$ originated from. The observed methylation status of each CpG may differ from the corresponding epiallele status with probability $\epsilon\in[0,1/2]$. Supposing $w_i=q$ we can therefore write $p(\vecy_i|\vecx_q,\epsilon,Q)=\prod_{\mu=1}^d p(y_{i\mu}|x_{q\mu},\epsilon,Q)$ where
\begin{equation}
p(y_{i\mu}|x_{q\mu},\epsilon,Q)=\left\{
\begin{array}{ll}
\epsilon & \quad\text{if $y_{i\mu}\neq x_{q\mu}$}\\
1-\epsilon & \quad\text{if $y_{i\mu}= x_{q\mu}$.}
\end{array}
\right.
\end{equation}
The epialleles are analogous to latent variables in a latent variable model. Our goal is to infer the quantities $\matX=(\vecx_1,\ldots,\vecx_Q)$ and $\vecw=(w_1,\ldots,w_N)$ as well as the hyperparameter $\epsilon$ and the number of epialleles $Q$ from the observed data $\matY=(\vecy_1,\ldots,\vecy_N)$. Using Bayes' theorem the posterior over the unknown quantities is
\begin{equation}
p(\matX,\vecw, \epsilon|\matY,Q) = \frac{p(\matY|\matX,\vecw,\epsilon,Q)p(\matX|Q)p(\vecw|Q)}{p(\matY|Q)}
\label{eq:posterior}
\end{equation}
where the likelihood is
\begin{align}
p(\matY|\matX,\vecw,\epsilon,Q) &=\prod_{i=1}^N\sum_{q=1}^Q \delta_{q,w_i}\,p(\vecy_i|\vecx_q,\epsilon,Q).
\label{eq:likelihood}
\end{align}
The delta function is defined by $\delta_{xy}=1$ if $x=y$ and $\delta_{xy}=0$ otherwise. The marginal density $p(\matY|Q)=\sum_{\matX'}\sum_{\vecw'}\int\text{d}\epsilon'\,p(\matY|\matX,\vecw,\epsilon,Q)p(\matX|Q)p(\vecw|Q)$ serves to normalise the posterior density where the summation is over all possible values of $\matX$ and $\vecw$. We will use maximum entropy priors which are uniform densities over the $2^{Qd}$ possible epiallele configurations $\matX$ and $Q^N$ possible values of $\vecw$. 

\subsection{Bayesian inference}

For fixed $\matX$, $\epsilon$, and $Q$, the maximum a posteriori (MAP) estimate for $\vecw$ is given by attributing each read $\vecy_i$ to the epiallele that is most similar to it. That is,
\begin{equation}
w_i^* = \text{argmax}_q  p(\vecy_i|\vecx_q,\epsilon,Q).
\label{eq:map_w}
\end{equation}
Next we wish to obtain the MAP estimate for $x_{q\mu}$ for fixed $\vecw$, $\epsilon$ and $Q$. Let $N_1$ denote the total number of methylated CpGs at site $\mu$ in observed reads that have been attributed to epiallele $q$. That is, $N_1 = \sum_i y_{i\mu}$ where the sum is restricted to indices for which $w_i=q$. Similarly, $N_0$ is the total number of unmethylated CpGs at site $\mu$ in reads stemming from epiallele $q$. It is straightforward to show that the MAP estimate is
\begin{align}
x^*_{q\mu} &= 1\quad\text{if $N_1 > N_0$}\nonumber\\
x^*_{q\mu} &= 0\quad\text{otherwise}\label{eq:map_x}.
\end{align}
We now define the total \emph{matches} at a given locus as $\alpha_1 = \sum_{i,\mu} \delta_{y_{i\mu},x_{w_i\mu}}$ and \emph{mismatches} as $\alpha_0 = \sum_{i,\mu} 1-\delta_{y_{i\mu},x_{w_i\mu}}$. It can be shown (see Supplementary Material) that the MAP estimate for $\epsilon$ is 
\begin{equation}
\epsilon^* = \frac{\alpha_0}{\alpha_0+\alpha_1}
\label{eq:map_e}
\end{equation}
which is simply the proportion of observed CpGs at that locus that differ from the underlying epialleles. Some values of $y_{i\mu}$ may be missing and these are handled by simply omitting them from sums and products over $i$ and $\mu$.

\subsubsection{Algorithm}

Note that the MAP estimates $\vecw^*$ and $\matX^*$ are independent of $\epsilon$. Given a set of observed data $\matY$ the first task is to determine optimal values for $\vecw$ and $\matX$. This is done according to the following algorithm:
\begin{enumerate}
\item
Initialise $\vecw$ by using hierarchical clustering to group the observed reads $\matY$ into $Q$ groups. The \emph{hamming distance} (the proportion of CpGs that differ between two sequencing reads) is used as a distance measure.
\item
Compute $\matX$ according to (\ref{eq:map_x}) using the current estimate of $\vecw$.
\item
Compute $\vecw$ according to (\ref{eq:map_w}) using the current estimate of $\matX$.
\item
Repeat steps 2 and 3 until $\vecw$ and $\matX$ converge to a steady solution (typically two or three iterations).
\end{enumerate}

Denote the final parameter values as $\hat{\vecw}$ and $\hat{\matX}$. The value for $\hat{\epsilon}$ is then given by (\ref{eq:map_e}).

\subsubsection{Model selection}

In principle, the marginal density $p(\matY|Q)$ could be used to compare models with different values of $Q$. In practice, however, $p(\matY|Q)$ is analytically intractable. Instead we use the Akaike information criterion (AIC) \citep{akaike1998information} in order to select the optimal number of epialleles
\begin{equation}
\text{AIC}(Q) = -2\log p(\matY|\hat{\matX},\hat{\vecw},\hat{\epsilon},Q) + 2Qd
\end{equation}
where $\hat{Q} = \text{argmin}_Q \text{AIC}(Q)$. For a model with $Q$ epialleles the $Qd$ parameters that make up the matrix $\matX$ are regarded as free parameters. The term $2Qd$ penalises more complex models (i.e. models with larger $Q$). A more complex model will only be selected if the evidence from the data is sufficiently strong to overcome the penalty term.

\subsubsection{Marginalisation of $\vecw$}

Finally, it may not be completely clear which epiallele an observed read should be attributed to (there could be several epialleles an equal edit distance away). This ambiguity manifests itself as the uncertainty surrounding the parameter $w_i$. The Bayesian approach allows this uncertainty to be incorporated into our analysis. The marginal density over $w_i$ is given by fixing all other parameters to their MAP values
\begin{equation}
p(w_i|\hat{\vecw}_{-i},\hat{\matX},\hat{\epsilon},\hat{Q}) = \frac{p(\matY|\hat{\matX},\hat{\vecw}_{-i},w_i,\hat{\epsilon},\hat{Q})p(\hat{\matX}|\hat{Q})p(\hat{\vecw}|\hat{Q})}{p(\matY|\hat{Q})}
\end{equation}
where $\hat{\vecw}_{-i}$ is a $(d-1)$-dimensional vector obtained from $\hat{\vecw}$ by removing element $i$. At the given locus in question the \emph{proportion of observed reads originating from epiallele $q$} is given by
\begin{equation}
\phi_q = \frac{1}{N}\sum_{i=1}^N p(w_i=q|\hat{\vecw}_{-i},\hat{\matX},\hat{\epsilon},\hat{Q}).
\label{eq:density}
\end{equation}
The quantity $\pv=(\phi_1,\ldots,\phi_{\hat{Q}})$ specifies the distribution of epialleles within that locus. An example is given in Figure \ref{fig:method} (d).

\subsection{Application to multi-region tumour sampling}

We will now describe our analysis protocol. In our application we are considering sequencing data from multiple regions of the same tumour. The number of distinct epialleles present at a particular locus is determined by pooling sequencing reads from all tissue samples (tumour and normal) in order to boost statistical power. Suppose there are $s=1,\ldots,S$ tumour samples with $N_s$ reads per sample (at a given locus). The total number of reads in the pool is now $N=\sum_s N_s$. Using the pooled reads a model is fitted as described above. The vector $\hat{\vecw}$ defines which epiallele each sequencing read originated from. The distribution of epialleles within region $s$ is given by
\begin{equation}
\phi_q^s = \frac{1}{N_s}\sum_{i\in I_s} p(w_i=q|\hat{\vecw}_{-i},\hat{\matX},\hat{\epsilon},\hat{Q})
\label{eq:density}
\end{equation}
where $I_s$ is the set of indices of reads belonging to sample $s$. The vectors $\pv^s$ serve to characterise each sample in terms of their epiallele distributions.

\subsubsection{Estimation of sample purity}
\label{sec:purity}

Suppose $\hat{Q}$ epialleles are inferred at particular locus of a particular tumour sample (for the sake of compactness we will not index the loci or samples). The locus is characterised by $\pv$, the inferred probability distribution over the $\hat{Q}$ epialleles. If the tumour sample is contaminated with normal tissue then we can write
\begin{equation}
\pv = \rho\vect + (1-\rho)\vecn
\label{eq:contamination}
\end{equation}
where $\rho\in[0,1]$ is the proportion of observed tissue that comes from the tumour (the sample `purity'), and $\vect$ and $\vecn$ are the epiallele distributions in the tumour and normal tissues respectively (at the particular locus in question). We can estimate $\pv$ and $\vecn$ from the observed data at a particular locus. Estimation of both $\rho$ and $\vect$ requires solving the $\hat{Q}$ equations in (\ref{eq:contamination}) for $\hat{Q}+1$ variables which generally is not possible. However, the quantity
\begin{equation}
\xi = \frac{1}{2}\sum_{\mu=1}^{\hat{Q}} \text{abs}(\phi_{\mu}-n_{\mu})
\label{eq:xi}
\end{equation}
can be computed at each locus of the observed tissue sample. If we substitute (\ref{eq:contamination}) into (\ref{eq:xi}) we can see that $\xi$ takes a minimum value of 0 when $\vect=\vecn$. At a locus in which the tumour and normal tissues have a completely different epiallele composition then we say that if $t_{\mu}>0$ then $n_{\mu}=0$ and if $n_{\mu}>0$ then $t_{\mu}=0$ for $\mu=1,\ldots,\hat{Q}$. It is straightforward to show that if this is the case then $\xi=\rho$ and that this is the maximum value $\xi$ can take.

We therefore expect that $\xi$ will take values in the range $[0,\rho]$ when computed across all loci of the observed sample. If we plot the empirical density of $\xi$ values the parameter $\rho$ can be estimated from the maximum value of $\xi$. Since $\pv$ and $\vecn$ are estimated from finite data samples we expect the distribution of $\xi$ to be `smoothed' by sampling noise. This is precisely what we observe in practice. An example of the empirical density of $\xi$ is plotted in Figure \ref{fig:purity}.

\subsubsection{Decontamination of normal tissue}

Finally, we note that once estimates of $\rho$ have been obtained we can calculate the `decontaminated' tumour epiallele profiles at each locus according to
\begin{equation}
\hat{t}_{\mu} = \frac{\phi_{\mu}-(1-\rho)n_{\mu}}{\rho}\quad\text{for $\mu=1,\ldots,\hat{Q}$.}
\end{equation}
We have used the notation $\hat{t}_{\mu}$ to emphasise that this is an estimate of the tumuor epiallele distribution. Due to the fact that $\pv$, $\vecn$ and $\rho$ are estimated from finite data samples it is possible that $\hat{t}_{\mu}$ can take values outside $[0,1]$. Any cases where $\hat{t}_{\mu}<0$ are set to 0 and any cases where $\hat{t}_{\mu}>1$ are set to 1.

A conventional anlaysis of DNAm sequencing data will typically `call' a methylation level at each CpG site by computing the proportion of reads on which a CpG is observed in a methylated state. Using our method a methylation level for each CpG site can readily be computed after decontamination of normal tissue and used in existing analysis pipelines.

\subsubsection{Construction of a phylogenetic tree}
\label{sec:tree}

Using the decontaminated representation of a sample $\hat{\vect}_s$ the euclidean distance between $\hat{\vect}_s$ and $\hat{\vect}_{s'}$ can be used as a distance measure between samples $s$ and $s'$. Each locus provides a distance matrix that depends on the distribution of epialleles at that particular locus. To obtain an overall distance matrix we average over distance matrices from all loci. Any distance based phylogenetic inference method can subsequently be used to construct a phylogenetic tree. We used the `fastme.bal' function as part of the `ape' R package \citep{paradis2004}.

%
%
\section{Results}
\label{sec:methods}
%

\begin{figure}
\centering
\includegraphics[scale=0.65]{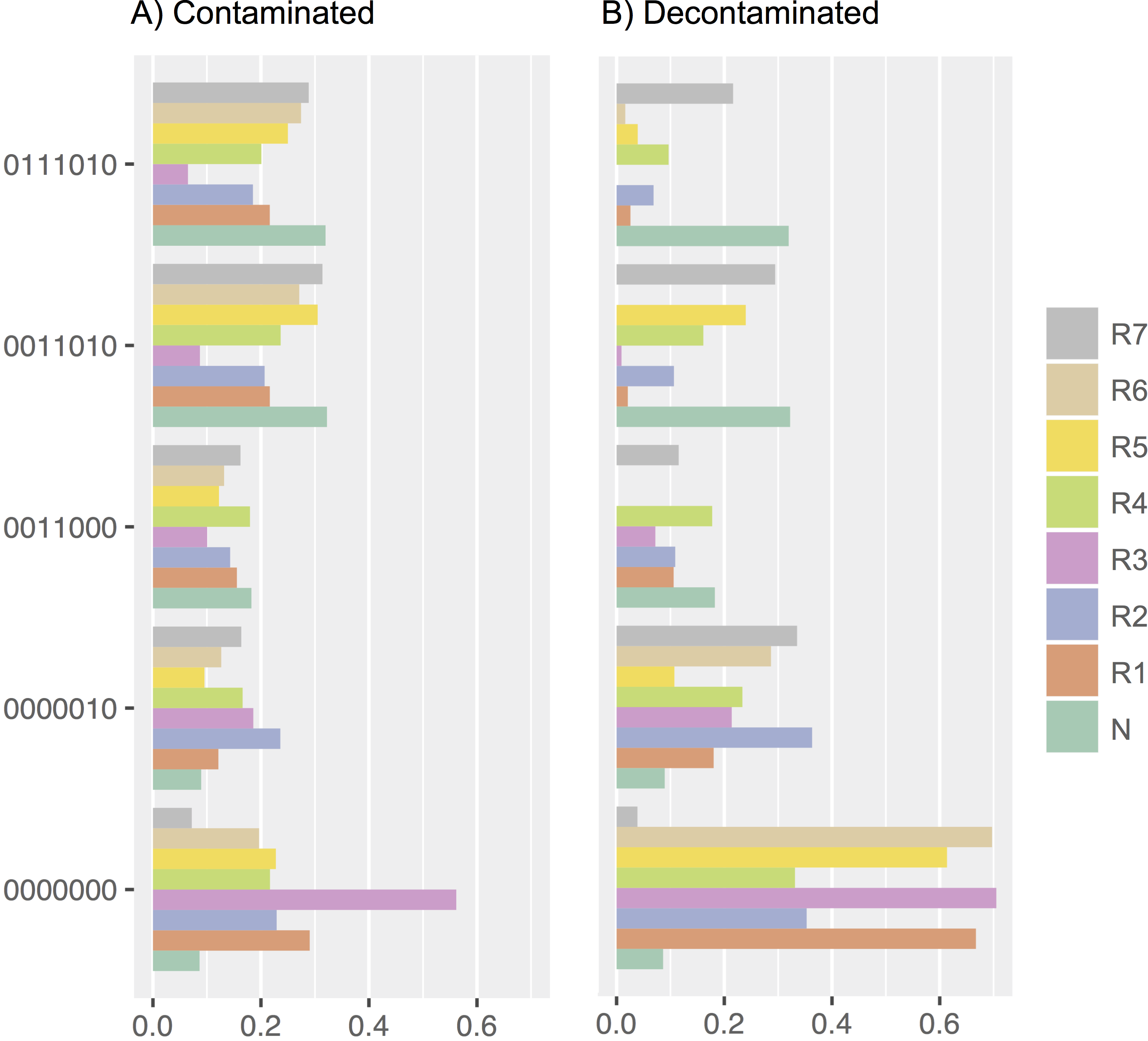}
\caption{A genomic locus (chr1:2,603,277-2,603,489) composed of seven CpGs. The distribution of five epialleles -- inferred using the Bayesian model -- are plotted for seven tumour regions (R1 to R7) and one normal sample (N). In (a) the tumour samples have not been corrected for normal tissue contamination whereas in (b) they have been. The tumour samples are shifting towards an unmethylated profile in comparison to the normal tissue.}
\label{fig:locus}
\end{figure}

\subsection{Simulations}

Simulations of a single locus were performed to study what effect the number of CpGs, $d$, the number of sequencing reads, $N$, and the noise level, $\epsilon$, have on our ability to correctly detect the underlying epialleles. The simulated reads were noise corrupted versions of three distinct randomly generated epialleles, and on average each epiallele corresponded to one third of the observed reads. To assess model performance we counted the proportion of observed reads that were attributed to their correct underlying epiallele (which requires both inference of the correct epialleles and attribution to the correct epiallele). For every value of the parameters results were averaged over 100 simulations.

We found that $N=100$ and $d=6$ gave a success rate of approximately 95\% at a 5\% noise level. These values were used to guide the selection of viable loci in subsequent analyses of experimental data. Dropping to $N=50$ gave a performance of just over 90\%  (Supplementary Figure 3). Sequencing depth beyond $N=100$ did not yield any additional performance gain. The performance saturates at 100\% for $d>15$ (Supplementary Figure 4). Since the number of possible epialleles is $2^d$ a larger $d$ will typically make it easier to resolve distinct epialleles. Additionally, since the underlying epialleles are randomly generated it is possible that some may be within one edit distance from each other, making it difficult for the model to distinguish between very similar epialleles and noise when $d$ is small. Performance was observed to decrease sharply for increasing noise levels (Supplementary Figure 5).

\subsection{Cell line data: detection of low frequency epialleles}

In order to test whether our statistical methods could detect low frequency epialleles in practice we mixed a fully unmethylated and fully methylated cell line in a 9:1 ratio prior to sequencing. Loci with six or more CpGs and 50 or more reads were identified. Within these loci 6.3\% of observed CpGs were methylated overall. The two cell lines were sequenced separately and we found that the fully methylated and unmethylated cells were in fact $97.3\%$ and $3.8\%$ methylated respectively.

The Bayesian model was used to detect the presence of epialleles at each loci. We found that 5.2\% of methylated CpGs were attributed to methylated epialleles (defined as epialleles with $\geq50\%$ methylation). The mean noise level was inferred as 1.1\%. This suggests that the majority of methylation is correctly identified as corresponding to a methylated profile and therefore our method is capable of resolving a distinct low frequency cellular subpopulation.

\begin{figure}[t]
\centering
\includegraphics[scale=0.75]{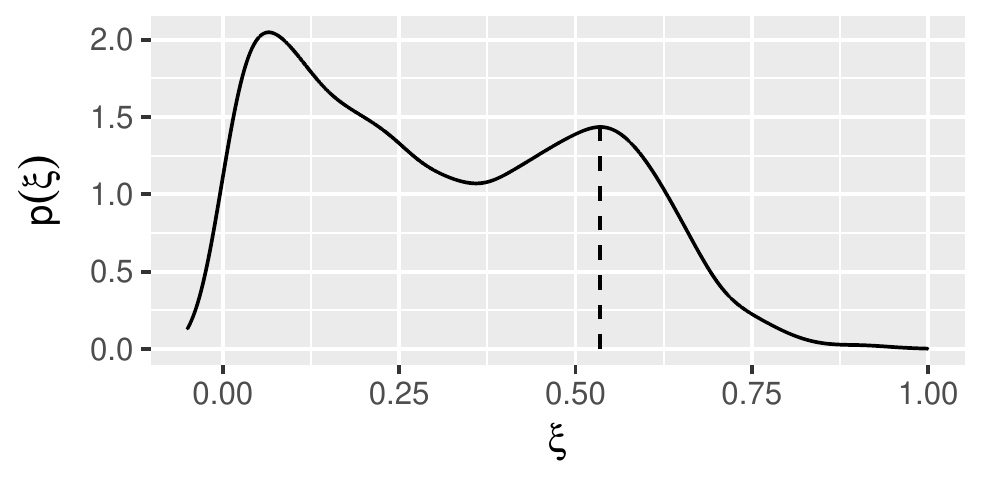}
\caption{Estimation of tumour sample purity for region 2 of the tumour. The parameter $\xi$ was calculated at all eligible loci across the genome and the empirical distribution is plotted here. The sample purity is equal to the maximum value of $\xi$ which is interpreted to occur at the rightmost maximum at $\xi=0.53$. The distribution of $\xi$ is `smoothed' due to the fact that at each locus $\xi$ is estimated from a finite sample of sequencing reads.}
\label{fig:purity}
\end{figure}

\subsection{Multi-region tumour sampling case study}

Our case study data consisted of seven tissue samples from a single lung tumour (CRUK00620) along with one matched normal tissue sample. These tissue samples were acquired as part of the larger TRACERx study \citep{nejm2017}. The raw sequencing data were trimmed and aligned to a reference genome. Sequencing reads were subsequently organised into distinct genomic loci as described in the Supplementary Material. We demanded that no more than 25\% of data were missing per locus (due to partially overlapping paired-end reads or reads not covering the whole locus). Any data from chromosomes X and Y were discarded. At each locus $\hat{Q}$ epialleles are inferred and any epialleles that accounted for less than 5\% of observed reads were discarded prior to the computation of $\pv_s$ for $s=1,\ldots,S$. This was done in order to focus on the dominant shifts in epiallele profiles and to minimise the risk of inferring spurious epialleles. 

In order to compare the distribution of epialleles within different tumour samples it was necessary to identify all of the loci which occurred in two or more samples. That is, the loci themselves must `match up' between tumour samples in order for a comparison to be made (partially overlapping loci were permitted provided they met the minimum number of non-missing CpG requirements). Only loci with a median read depth $\geq100$ across normal and tumour tissue samples and six or more CpGs were considered. A total of 39,940 loci were analysed out of which 73\% were found to contain a single epiallele, 13\% contained two, 7\% contained three, 4\% contained four, and 3\% had five our more (up to a maximum of thirteen).

\begin{figure*}
\centering
\includegraphics[scale=0.58]{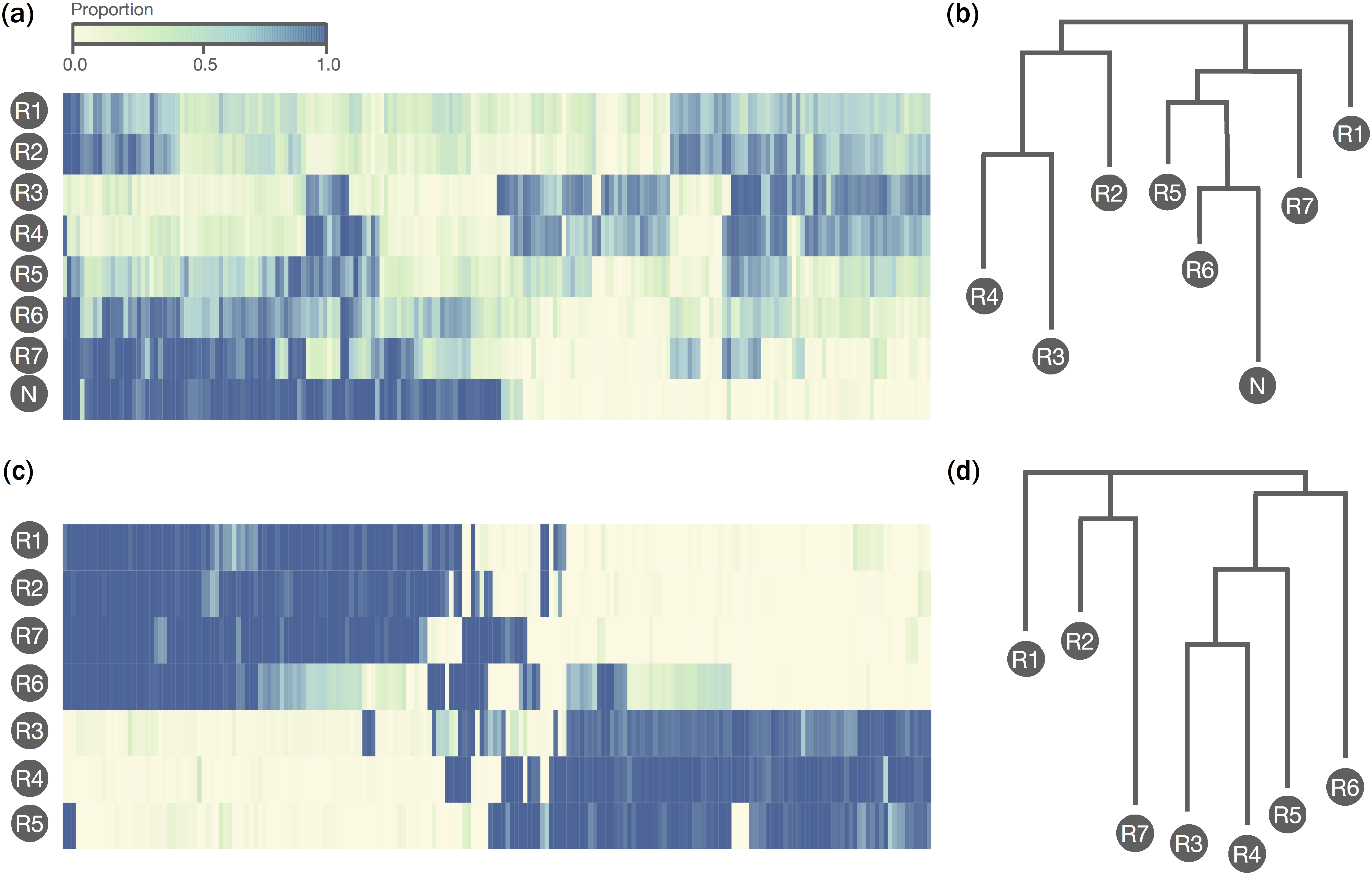}
\caption{(a) Heatmap of the top 200 most variable epialleles across the seven tumour samples (labelled R1 to R7) and matched normal sample (labelled N). A proportion of 1.0 (dark blue) means that that epiallele accounted for all observed methylation patterns at the corresponding locus. These data have not been decontaminated of normal tissue. (b) The phylogenetic tree inferred before correction for contaminating normal tissue. In (c) and (d) are the same figures for the decontaminated epiallele profiles.}
\label{fig:heatmap}
\end{figure*}

\subsubsection{Comparison of epiallele distribution throughout the tumour}

At each locus the Bayesian model is used to infer the epialleles present, the total number of epialleles, and which epialleles each observed sequence came from. An example locus with seven CpGs from chromosome one is presented in Figure \ref{fig:locus}. At this locus five distinct epialleles were detected. Both the observed and decontaminated profiles are shown. The normal tissue is predominantly composed of methylated epialleles whereas the tumour samples have a greater proportion of less methylated epialleles. This suggests that within the tumour there exist cellular subpopulations that are undergoing a transition from a methylated state to an unmethylated one.

In order to understand shifts in epiallele frequency at a global level we plotted a heatmap of the top 200 most variable epialleles in Figure \ref{fig:heatmap} (a) and (c). Both the observed and decontaminated epiallele profiles were used. Tumour samples are characterised by both a loss and gain of numerous epialleles when compared to the normal tissue sample. The variability in epiallele expression throughout different parts of the tumours suggests that a substantial level of tumour heterogeneity exists at the epigenetic level. Note that in the contaminated samples 71 out of the 200 epialleles were located on CpG islands, and 54 were located on a CpG shore (defined as 2 kilobases either side of an island). In the decontaminated version 124 epialleles were located on an island and 38 on a shore.

\subsubsection{Estimation of sample purity}

The sample purities were estimated as described in Section \ref{sec:purity}. An example of the empirical density of $\xi$ within tumour region 2 is plotted in Figure \ref{fig:purity}. From the location of the rightmost maximum we estimate $\rho=0.535$. Plots for all tumour regions are given in Supplementary Figure 6. Estimates of purity for the seven tumour samples are given in Table \ref{tab:purity}. For tumour region 6 the rightmost maxima was not visible presumably due to very low tumour purity. The purity estimates are compared to estimates obtained from an analysis of exome data from the same tissue samples performed independently in \citet{nejm2017}.

\begin{table}[b!]
\centering
\begin{tabular}{|c|c|c|}
\hline
Tumour sample & Epiallele purity estimate & Exome purity estimate\\
\hline
R1 & 35\% & 32\%\\
R2 & 54\% & 51\%\\
R3 & 75\% & 73\%\\
R4 & 53\% & 67\%\\
R5 & 25\% & 28\%\\
R6 & <20\% & 13\%\\
R7 & 30\% & 36\%\\
\hline
\end{tabular}
\caption{In the middle column are estimates of tumour purity based on a comparison of epiallele distributions between normal tissue and tumour tissue. The third column contains estimates obtained from a separate study of exome data from the same tumour samples.}
\label{tab:purity}
\end{table}

\begin{figure}[t]
\centering
\includegraphics[scale=0.55]{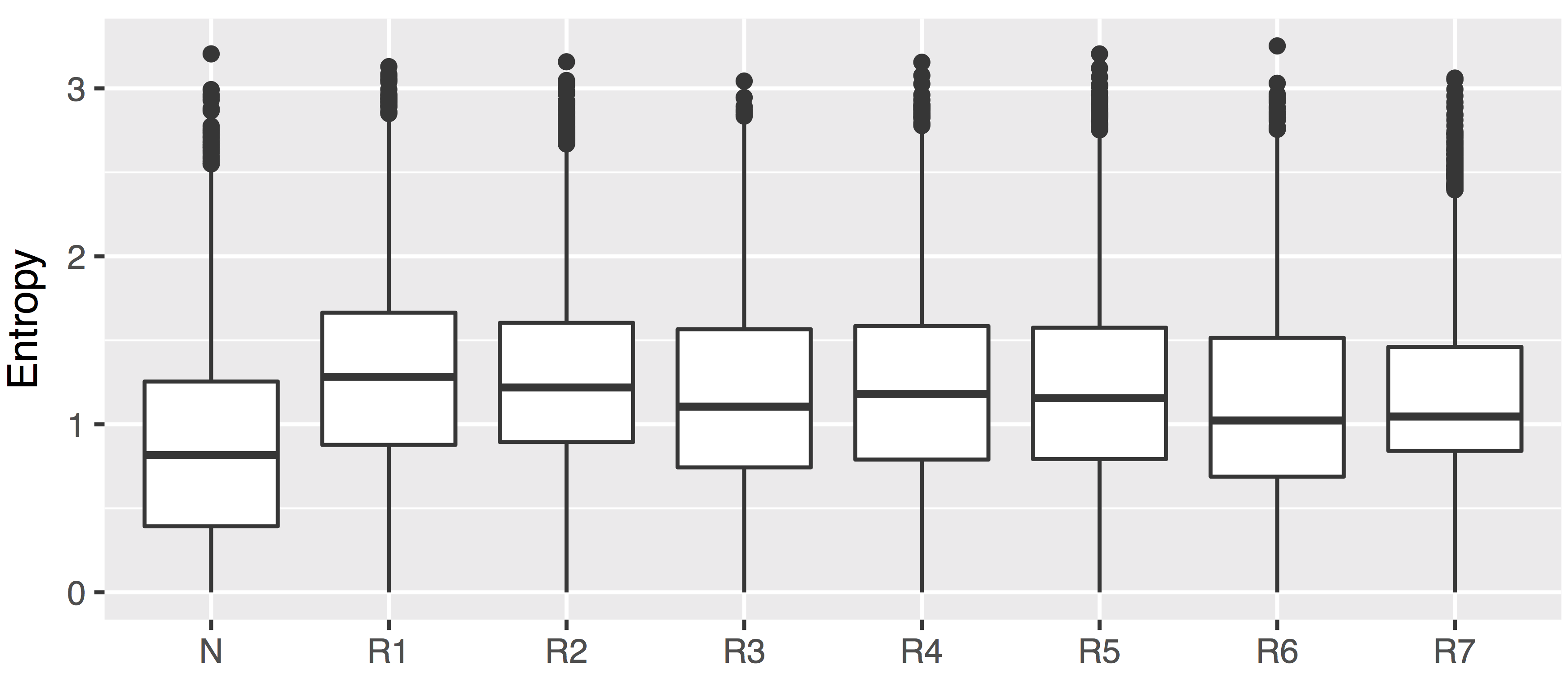}
\caption{Box plots of the Shannon entropy of the epiallele distribution across normal tissue (N) and the seven tumour regions (R1--R7).}
\label{fig:entropy}
\end{figure}

\subsubsection{Inference of a phylogenetic tree}

Phylogenetic trees were generated as described in Section \ref{sec:tree}. The trees for both contaminated and decontaminated samples are plotted in Figure \ref{fig:heatmap} (b) and (d). The structure of the contaminated tree is dominated by the sample purities, with low purity samples clustering together. The decontaminated tree has a totally different structure and this is broadly similar to a phylogenetic tree obtained from from a separate genetic analysis of the same patient and shown in Supplementary Figure 7.

\subsubsection{Quantification of epigenetic disorder}

The Shannon entropy provides a measure of how disordered a random variable is. In particular, the entropy of the epiallele distribution $\pv_s$ quantifies how disordered or heterogeneous each loci is in sample $s$. In Figure \ref{fig:entropy} box plots summarise the distribution of entropies across tumour and normal tissues (without decontamination). The tumour tissue samples have a substantially elevated entropy in comparison to the normal tissue.

%
%
\section{Discussion}
\label{sec:discussion}
%
%

Analysis of epialleles allows for a deeper interrogation of the underlying biology than a pointwise examination of CpG methylation states. Tracing the patterns of DNA methylation along epialleles allows one to tease apart different cellular subpopulations and acquire a richer quantification of heterogeneity and disorder that would not be possible by looking at individual CpG sites. In particular, the distribution of epialleles throughout a tumour can shed light on the evolutionary history of the tumour.

Our analysis protocol specifically pools sequencing reads from multiple tissue samples in order to leverage greater statistical power in epiallele detection. Our Bayesian approach will automatically detect the number of epialleles present, and infer what the methylation pattern of those epialleles are. One strength of the Bayesian approach is that it provides a framework for averaging over uncertainty in model parameters. If there is uncertainty as to which epiallele an observed sequencing read may have originated from, then a natural solution is to average over that uncertainty by marginalising over the appropriate posterior distribution. In addition to the above features our model can easily accommodate missing data and can handle an arbitrary sequencing depth and number of CpG sites per locus. Furthermore, by comparing the distribution of epialleles within normal and tumour tissue samples it is possible to estimate the purity of each sample and to subsequently decontaminate them. Methylation levels at each CpG site can be extracted from the decontaminated samples and subsequently used in standard analysis pipelines.

Tracking the presence or absence of epialleles throughout the tumour opens up an additional layer of complexity beyond that of conventional methylation analyses. Pointwise methylation analysis protocols typically average over sequencing reads -- to `call' the methylation status at single CpGs -- that potentially come from a diverse and heterogenous population of cells. Detecting which epialleles are present allows one to distinguish between these cellular subpopulations and identify tumour subclones that are defined by distinct epialleles. One can then probe changes between normal and cancerous tissue at a finer resolution. As we have demonstrated here, studying epiallele frequencies in different parts of the tumour reveals the evolutionary history of the tumour and allows a phylogenetic tree to be constructed. A measure of disorder or heterogeneity inside the tumour can be obtained through measures such as Shannon's entropy.

%
%
\section{Conclusion}
\label{sec:conclusion}
%
%

Understanding tumour heterogeneity is an important step towards understanding why certain therapies fail and why resistance to treatment can emerge. Subclonal populations of treatment-resistant cells can persist after treatment even if they only account for a small fraction of the original tumour. Epigenetic diversity within the tumour may play an important role in tumour evolution alongside genetic variability. It is increasingly recognised that for DNA methylation sequencing data studying the patterns of methylation along the genome -- `epialleles' -- can provide greater insight into the underlying dynamics of epigenetic regulation than a conventional pointwise analysis.

We have exploited this opportunity to study the distribution of epialleles throughout a tumour by performing reduced representation bisulfite sequencing on seven regions of the same tumour and one matched normal tissue sample. Our new Bayesian approach infers which epialleles are present at a given locus. A comparison of the frequency of different epialleles across the tumour and normal tissue highlights changes between normal and cancerous tissue and allows the extraction of a phylogenetic history. The concept of entropy can be used as a measure of global disorder within the tumour. Our method can be applied more generally to any type of DNAm sequencing data.

Future work will focus on larger scale studies of multiple patients with multi-region tumour sampling in order to probe for systematic alterations in epiallele expression between normal and cancerous tissue. Previously, measures of epigenetic disorder were found to be associated with clinical outcome and it will be interesting to see if quantification of disorder at the level of epialleles will provide a more refined measure of tumour aggressiveness. Ultimately, it is hoped that a clearer elucidation of epigenetic dynamics will complement our genetic knowledge of cancer and provide a more comprehensive understanding of the disease.

%
%
\section*{Acknowledgments}
%
%

The authors would like to thank Pawan Dhami (UCL Cancer Institute Genomics Core Facility) for sequencing support.

%
%
\section*{Funding}
%
%

JB was supported by the CRUK \& EPSRC Comprehensive Cancer Imaging Centre at King's College London and University College London jointly funded by Cancer Research UK and the EPSRC; AF by the MRC (MR/M025411/1); JH by the UCL Cancer Institute Research Trust; MT by the People Programme (Marie Curie Actions) of the EU Seventh Framework Programme (FP7/2007-2013/608765) and the Danish Council for Strategic Research (1309-00006B); GAW is funded by Cancer Research UK (grant number C11496/A17786); SB by NIHR-BRC (BRC275/CN/SB/101330) and the Wellcome Trust (99148); CS is Royal Society Napier Research Professor; This work was supported by the Francis Crick Institute which receives its core funding from Cancer Research UK (FCI01), the UK Medical Research Council (FC001169), and the Wellcome Trust (FC001169); by the UK Medical Research Council (MR/FC001169/1); CS is funded by Cancer Research UK (TRACERx), the CRUK Lung Cancer Centre of Excellence, Stand Up 2 Cancer (SU2C), the Rosetrees Trust, NovoNordisk Foundation (ID 16584), the Prostate Cancer Foundation, the Breast Cancer Research Foundation (BCRF), the European Research Council (THESEUS) and Marie Curie Network PloidyNet. Support was also provided to CS by the National Institute for Health Research, the University College London Hospitals Biomedical Research Centre, and the Cancer Research UK University College London Experimental Cancer Medicine Centre.

\bibliographystyle{plainnat}
\bibliography{refs}

\begin{thebibliography}{21}
\providecommand{\natexlab}[1]{#1}
\providecommand{\url}[1]{\texttt{#1}}
\expandafter\ifx\csname urlstyle\endcsname\relax
  \providecommand{\doi}[1]{doi: #1}\else
  \providecommand{\doi}{doi: \begingroup \urlstyle{rm}\Url}\fi

\bibitem[Akaike(1998)]{akaike1998information}
H~Akaike.
\newblock Information theory and an extension of the maximum likelihood
  principle.
\newblock In \emph{Selected Papers of Hirotugu Akaike}, pages 199--213.
  Springer, 1998.

\bibitem[Brocks et~al.(2014)Brocks, Assenov, Minner, Bogatyrova, Simon, Koop,
  Oakes, Zucknick, Lipka, Weischenfeldt, et~al.]{brocks2014intratumor}
David Brocks, Yassen Assenov, Sarah Minner, Olga Bogatyrova, Ronald Simon,
  Christina Koop, Christopher Oakes, Manuela Zucknick, Daniel~Bernhard Lipka,
  Joachim Weischenfeldt, et~al.
\newblock Intratumor dna methylation heterogeneity reflects clonal evolution in
  aggressive prostate cancer.
\newblock \emph{Cell Rep.}, 8\penalty0 (3):\penalty0 798--806, 2014.

\bibitem[Fang et~al.(2012)]{fang2012genomic}
F.~Fang et~al.
\newblock {Genomic landscape of human allele-specific DNA methylation}.
\newblock \emph{Proc. Natl. Acad. Sci.}, 109\penalty0 (19):\penalty0
  7332--7337, 2012.

\bibitem[Gu et~al.(2011)Gu, Smith, Bock, Boyle, Gnirke, and
  Meissner]{gu2011preparation}
Hongcang Gu, Zachary~D Smith, Christoph Bock, Patrick Boyle, Andreas Gnirke,
  and Alexander Meissner.
\newblock Preparation of reduced representation bisulfite sequencing libraries
  for genome-scale dna methylation profiling.
\newblock \emph{Nat. Protoc.}, 6\penalty0 (4):\penalty0 468--481, 2011.

\bibitem[He et~al.(2013)]{he2013dmeas}
J.~He et~al.
\newblock {DMEAS}: {DNA} methylation entropy analysis software.
\newblock \emph{Bioinformatics}, 2013.

\bibitem[{Jamal-Hanjani et al.}(2017)]{nejm2017}
{Jamal-Hanjani et al.}
\newblock {TRACERx --- Tracking Non-Small Cell Lung Cancer Evolution}.
\newblock \emph{Manuscript under revision}, 2017.

\bibitem[Landan et~al.(2012)]{landan2012epigenetic}
G.~Landan et~al.
\newblock {Epigenetic polymorphism and the stochastic formation of
  differentially methylated regions in normal and cancerous tissues}.
\newblock \emph{Nat. Genet.}, 44\penalty0 (11):\penalty0 1207--1214, 2012.

\bibitem[Landau et~al.(2014)]{landau2014locally}
D.~Landau et~al.
\newblock Locally disordered methylation forms the basis of intratumor
  methylome variation in chronic lymphocytic leukemia.
\newblock \emph{Cancer Cell}, 26\penalty0 (6):\penalty0 813--825, 2014.

\bibitem[Lee et~al.(2015)]{lee2015new}
S.~Lee et~al.
\newblock {New approaches to identify cancer heterogeneity in DNA methylation
  studies using the {L}epage test and multinomial logistic regression}.
\newblock In \emph{Computational Intelligence in Bioinformatics and
  Computational Biology (CIBCB), 2015 IEEE Conference on}, pages 1--7, 2015.

\bibitem[Li et~al.(2014)]{li2014dynamic}
S.~Li et~al.
\newblock Dynamic evolution of clonal epialleles revealed by methclone.
\newblock \emph{Genome Biol.}, 15\penalty0 (9):\penalty0 1, 2014.

\bibitem[Lin et~al.(2015)]{lin2015estimation}
P.~Lin et~al.
\newblock Estimation of the methylation pattern distribution from deep
  sequencing data.
\newblock \emph{BMC Bioinform.}, 16\penalty0 (1):\penalty0 1, 2015.

\bibitem[Mazor et~al.(2016)]{mazor2016intratumoral}
T.~Mazor et~al.
\newblock Intratumoral heterogeneity of the epigenome.
\newblock \emph{Cancer cell}, 29\penalty0 (4):\penalty0 440--451, 2016.

\bibitem[Pan et~al.(2015)Pan, Jiang, Boi, Tabb{\`o}, Redmond, Nie, Ladetto,
  Chiappella, Cerchietti, Shaknovich, et~al.]{pan2015epigenomic}
Heng Pan, Yanwen Jiang, Michela Boi, Fabrizio Tabb{\`o}, David Redmond, Kui
  Nie, Marco Ladetto, Annalisa Chiappella, Leandro Cerchietti, Rita Shaknovich,
  et~al.
\newblock Epigenomic evolution in diffuse large b-cell lymphomas.
\newblock \emph{Nat. Commun.}, 6, 2015.

\bibitem[Paradis et~al.(2004)]{paradis2004}
E.~Paradis et~al.
\newblock A{PE}: analyses of phylogenetics and evolution in {R} language.
\newblock \emph{Bioinformatics}, 20:\penalty0 289--290, 2004.

\bibitem[Peng and Ecker(2012)]{peng2012detection}
Q.~Peng and J.~Ecker.
\newblock Detection of allele-specific methylation through a generalized
  heterogeneous epigenome model.
\newblock \emph{Bioinformatics}, 28\penalty0 (12):\penalty0 i163--i171, 2012.

\bibitem[Richards(2006)]{richards2006inherited}
E.~Richards.
\newblock Inherited epigenetic variation---revisiting soft inheritance.
\newblock \emph{Nat. Rev. Genet.}, 7\penalty0 (5):\penalty0 395--401, 2006.

\bibitem[Sheffield et~al.(2017)Sheffield, Pierron, Klughammer, Datlinger,
  Sch{\"o}negger, Schuster, Hadler, Surdez, Guillemot, Lapouble,
  et~al.]{sheffield2017dna}
Nathan~C Sheffield, Gaelle Pierron, Johanna Klughammer, Paul Datlinger, Andreas
  Sch{\"o}negger, Michael Schuster, Johanna Hadler, Didier Surdez, Delphine
  Guillemot, Eve Lapouble, et~al.
\newblock Dna methylation heterogeneity defines a disease spectrum in ewing
  sarcoma.
\newblock \emph{Nat. Med.}, 2017.

\bibitem[Suzuki and Bird(2008)]{suzuki2008dna}
M.~Suzuki and A.~Bird.
\newblock Dna methylation landscapes: provocative insights from epigenomics.
\newblock \emph{Nat. Rev. Genet.}, 9\penalty0 (6):\penalty0 465--476, 2008.

\bibitem[Wu et~al.(2015)]{wu2015nonparametric}
X.~Wu et~al.
\newblock Nonparametric bayesian clustering to detect bipolar methylated
  genomic loci.
\newblock \emph{BMC Bioinform.}, 16\penalty0 (1):\penalty0 1, 2015.

\bibitem[Xie et~al.(2011)]{xie2011genome}
H.~Xie et~al.
\newblock {Genome-wide quantitative assessment of variation in DNA methylation
  patterns}.
\newblock \emph{Nucleic Acids Res.}, 39\penalty0 (10):\penalty0 4099--4108,
  2011.

\bibitem[Zheng et~al.(2014)]{zheng2014methylpurify}
X.~Zheng et~al.
\newblock {MethylPurify: tumor purity deconvolution and differential
  methylation detection from single tumor DNA methylomes}.
\newblock \emph{Genome Biol.}, 15\penalty0 (7):\penalty0 1, 2014.

\end{thebibliography}

%
%
\newpage
\appendix
%
%

\section{Extraction of viable loci}

\begin{figure}[h!]
\centering
\includegraphics[scale=0.6]{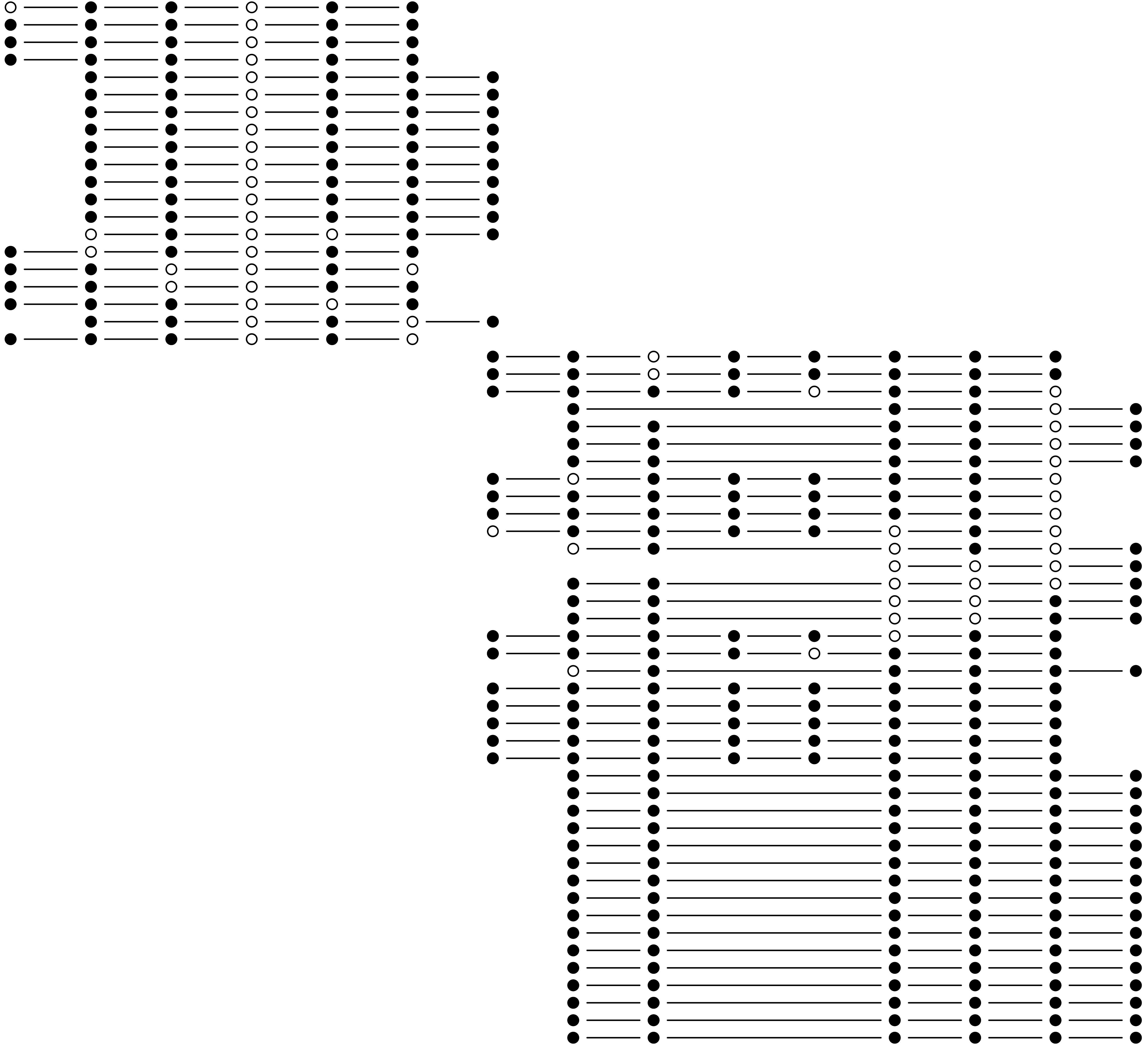}
\caption{An example of an observed locus (chr1:15,232,224-15,232,587) before preprocessing.}
\label{fig:example-locus}
\end{figure}

In Figure \ref{fig:example-locus} is an example of an observed locus before any preprocessing steps have been taken. The locus is defined naturally since it does not overlap with any other observed sequencing reads. Note that several sequencing reads only partially overlap and many do not overlap at all. In addition, several CpG measurements are missing from the middle of some reads due to the paired-end sequencing protocol that was used to generate the data (sometimes the paired ends may not span the full length of the DNA fragment).

Intuitively this locus should be split into two loci as the two blocks of reads overlap by a single CpG which is not enough to phase the inferred epialleles. We implemented the following algorithm to split the observed sequencing reads into sensible loci.

\begin{enumerate}
\item
Specify the minimum number of contiguous CpGs $d_{min}$ and the minimum number of reads $N_{min}$ required in order for a locus to be admissible (we used $d_{min}=6$ and $N_{min}=100$ in practice).
\item
Specify the maximum proportion of missing data that is allowed (we used 25\% in practice).
\item
If the observed locus contains more than 25\% missing data move to the next step, otherwise skip to step 4.
\begin{itemize}
\item[3(a).]
Represent the observed reads as a matrix with all non-missing measurements equal to 1 and all missing values represented with 0. Use hierarchical clustering to split the reads into two groups using the hamming distance.
\item[3(b).]
If the two groups contain less than 25\% of the missing data then proceed to the next step. Otherwise repeat the clustering above with three groups and so forth.
\end{itemize}
\item
Discard any loci that fail to meet the minimum values of $d$ and $N$.
\end{enumerate}

\begin{figure}[h!]
\centering
\begin{tabular}{l r}
\subfloat{\includegraphics[scale=0.6]{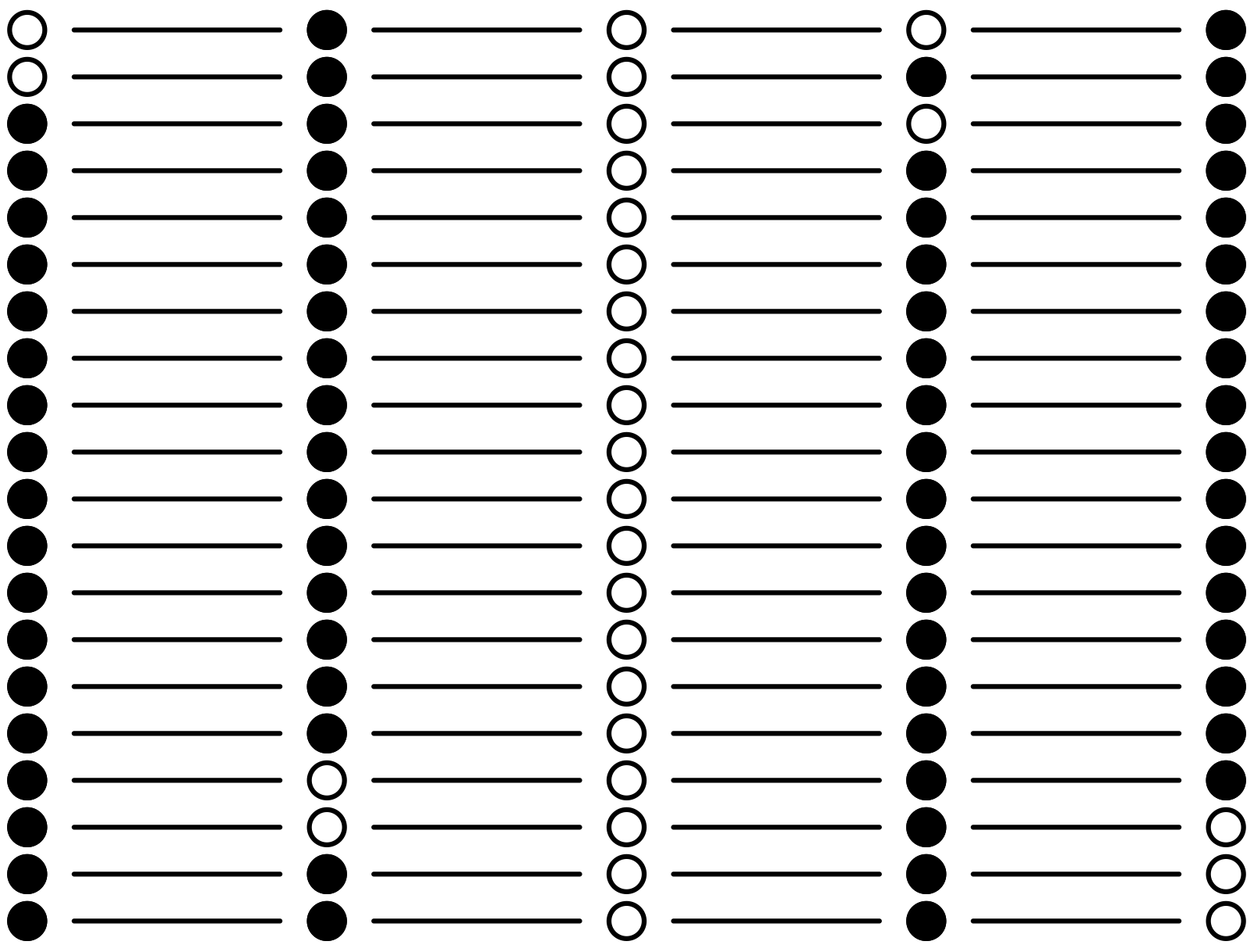}} & \subfloat{\includegraphics[scale=0.6]{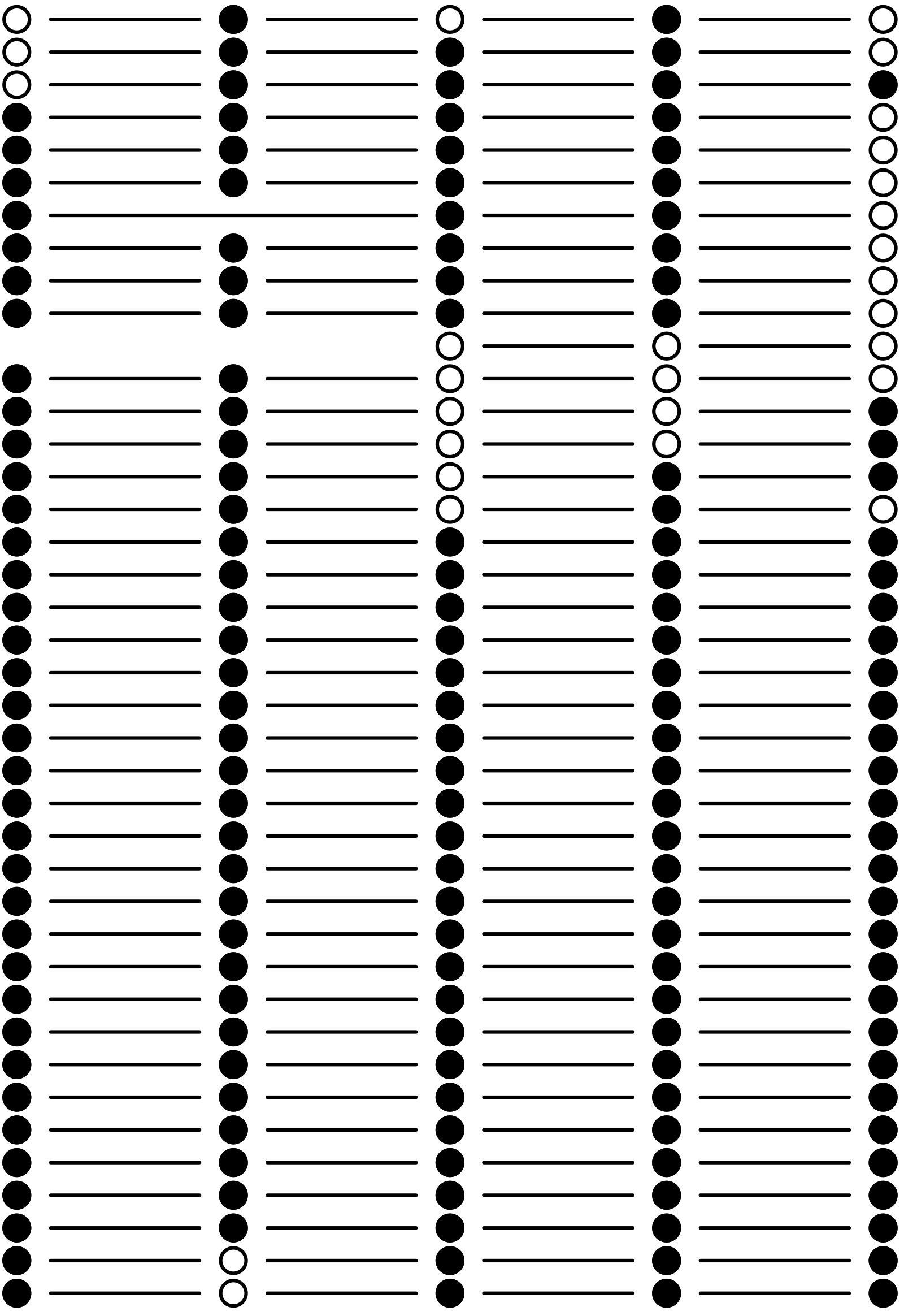}}\\
\end{tabular}
\caption{The two loci that are extracted from the observed data in Figure \ref{fig:example}.}
\label{fig:example2}
\end{figure}

We demanded that, in addition to the loci as a whole, each CpG site should not contain more than 25\% missing data. Any CpGs that failed to meet this constraint were discarded. This helped to trim low-coverage CpGs from at the edges of observed loci (for example, the very leftmost CpG in Figure \ref{fig:example}). The resulting loci after preprocessing are shown in Figure \ref{fig:example2}.

\newpage
\section{MAP estimate for $\epsilon$}

Given $\matX$, $\vecw$, $Q$ and uniform priors $p(\matX|Q)$ and $p(\vecw|Q)$ the MAP estimate for the hyperparameter $\epsilon$ is given by maximising the log of the posterior (2):
\begin{align}
\mathcal{L}(\epsilon) &= \log p(\matY|\matX,\vecw,\epsilon,Q)\nonumber\\
&=\log \prod_{i=1}^N \prod_{\mu=1}^d p(y_{i\mu}|x_{w_i\mu},\epsilon,Q)\nonumber\\
&=\alpha_0\log\epsilon + \alpha_1\log(1-\epsilon).
\end{align}
Note that on the second line that if read $\vecy_i$ originates from epiallele $q$ then $w_i=q$. Recall that  $\alpha_1 = \sum_{i,\mu} \delta_{y_{i\mu},x_{w_i\mu}}$ and  $\alpha_0 = \sum_{i,\mu} 1-\delta_{y_{i\mu},x_{w_i\mu}}$ denote the total number of matches and mismatches between the observed reads $\vecy$ and the corresponding epialleles $\vecx$ at this particular loci. Solving $\text{d}\mathcal{L}/\text{d}\epsilon = 0$ yields $\epsilon=\alpha_0/(\alpha_0+\alpha_1)$.

\newpage
\section{Simulation results}

\begin{figure}[h!]
\centering
\includegraphics[scale=0.6]{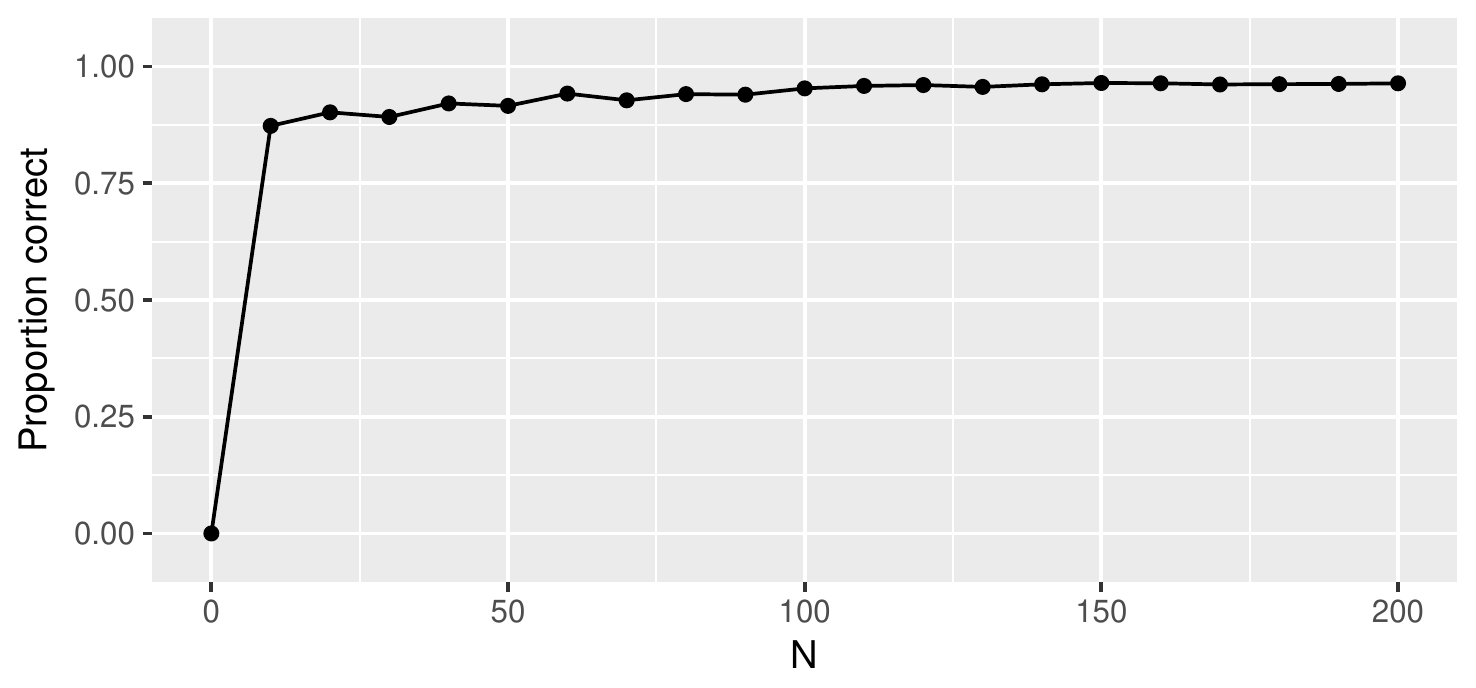}
\caption{Proportion of observed reads attributed to the correct underlying epiallele as a function of $N$ (the number of sequencing reads at the simulated locus). Parameters were fixed to $\epsilon=0.05$, $d=6$ and $Q=3$.}
\label{fig:simN}
\end{figure}

\begin{figure}[h!]
\centering
\includegraphics[scale=0.6]{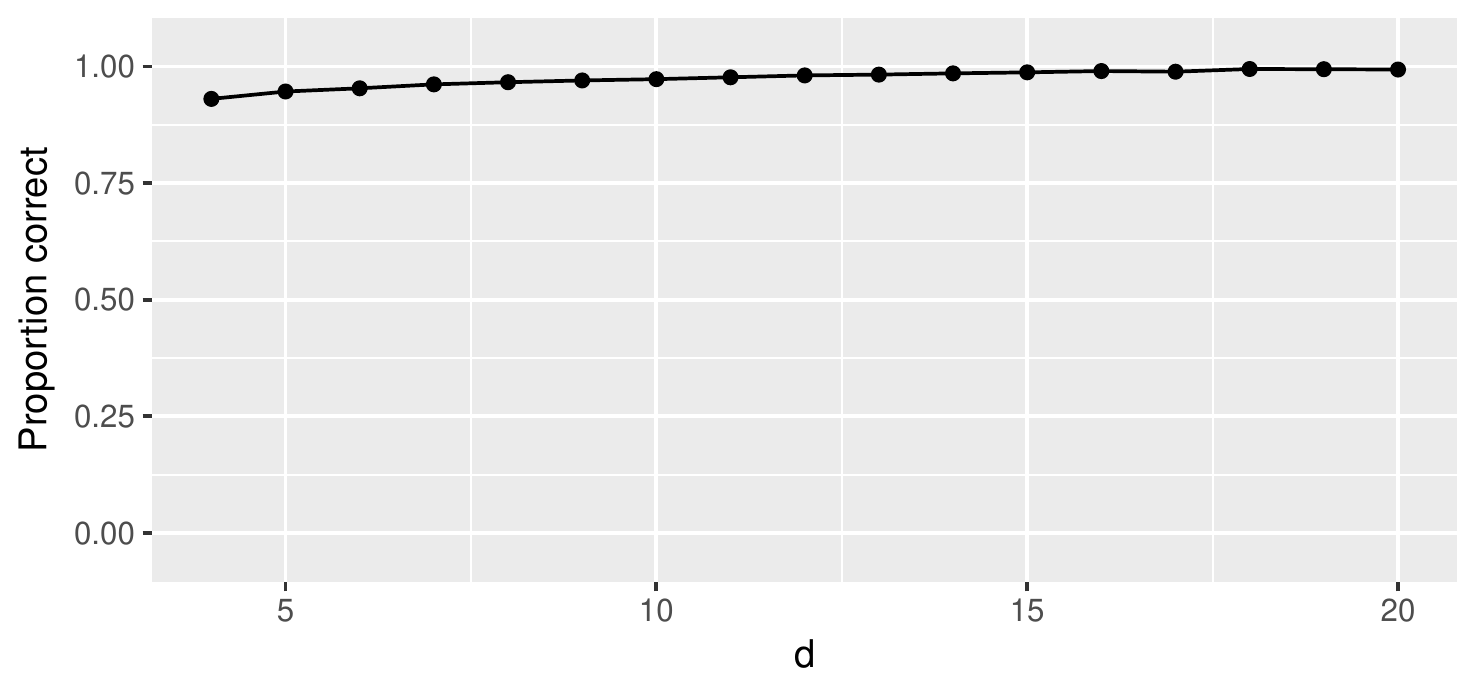}
\caption{Proportion of observed reads attributed to the correct underlying epiallele as a function of $d$ (the number of CpGs at the simulated locus). Parameters were fixed to $N=100$, $\epsilon=0.05$ and $Q=3$.}
\label{fig:simd}
\end{figure}

\begin{figure}[h!]
\centering
\includegraphics[scale=0.6]{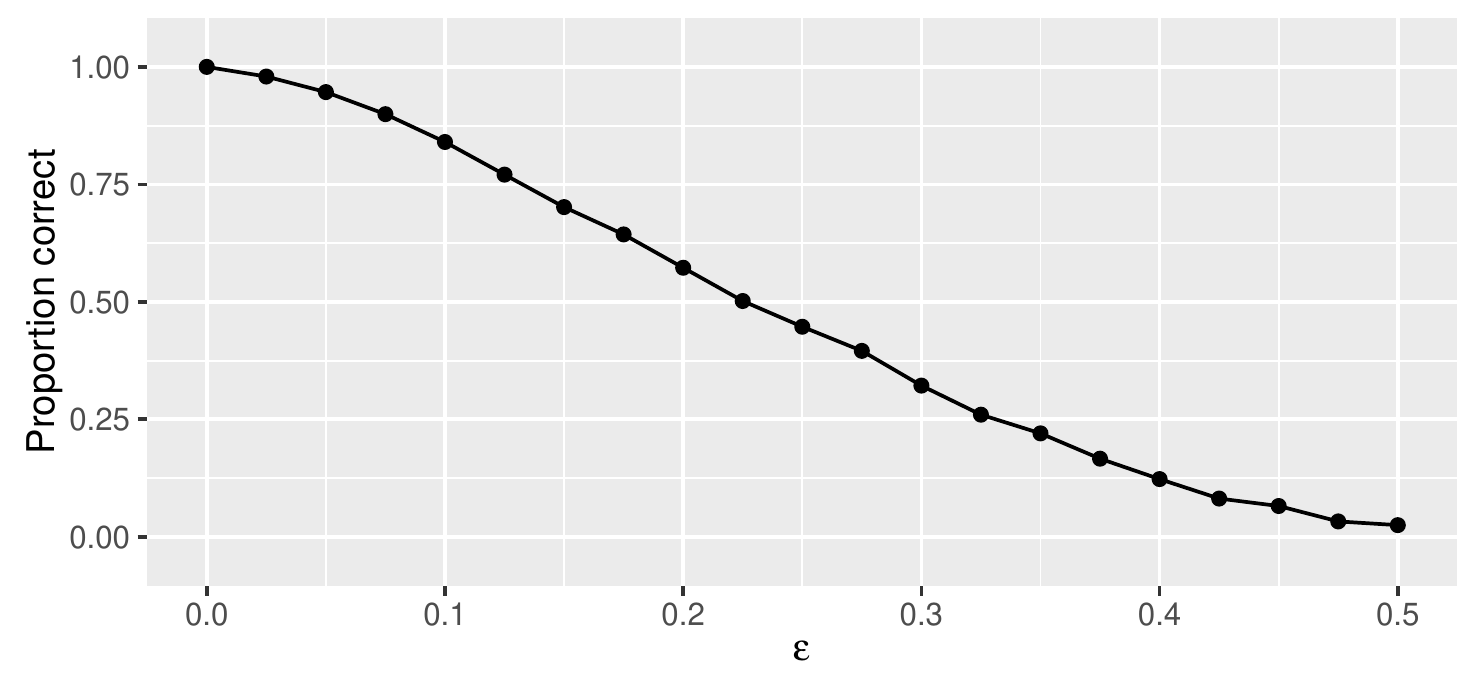}
\caption{Proportion of observed reads attributed to the correct underlying epiallele as a function of $\epsilon$ (the noise level). Parameters were fixed to $N=100$, $d=6$ and $Q=3$.}
\label{fig:simbeta}
\end{figure}

\newpage
\section{Purity estimation}
\begin{figure}[h!]
\centering
\begin{tabular}{c c}
\subfloat[R1]{\includegraphics[scale=0.6]{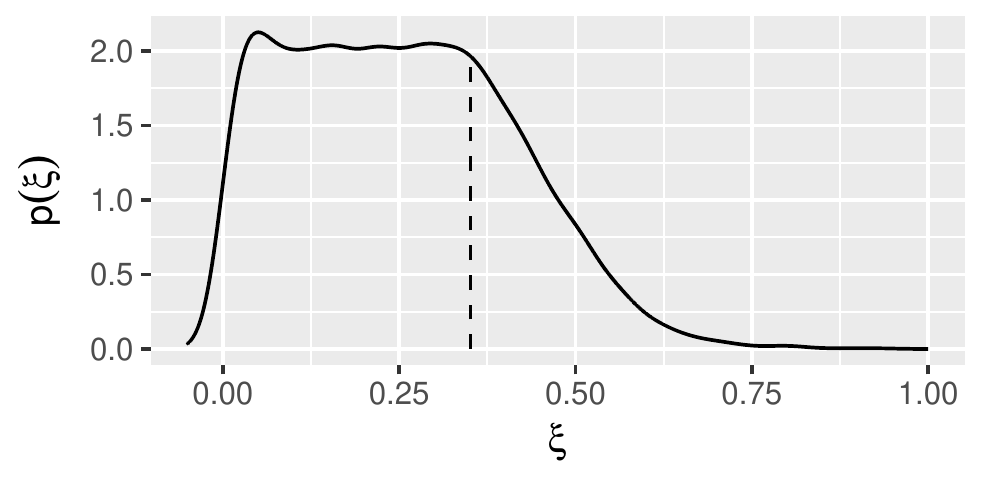}} & \subfloat[R2]{\includegraphics[scale=0.6]{density_R2.pdf}}\\
\subfloat[R3]{\includegraphics[scale=0.6]{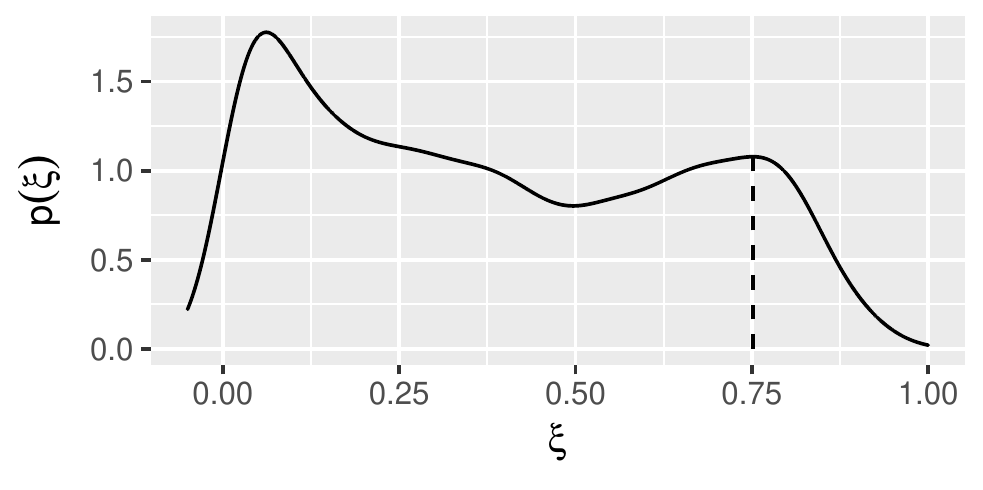}} & \subfloat[R4]{\includegraphics[scale=0.6]{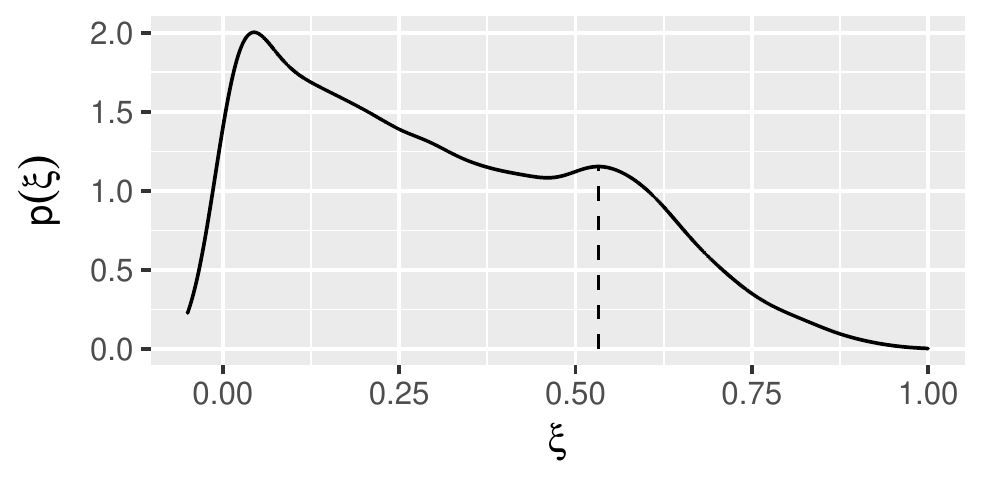}}\\
\subfloat[R5]{\includegraphics[scale=0.6]{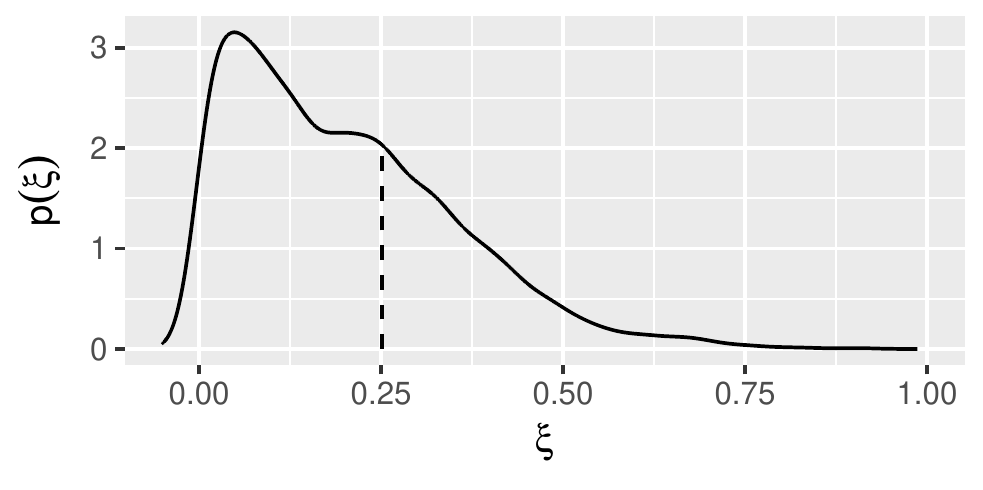}} & \subfloat[R6]{\includegraphics[scale=0.6]{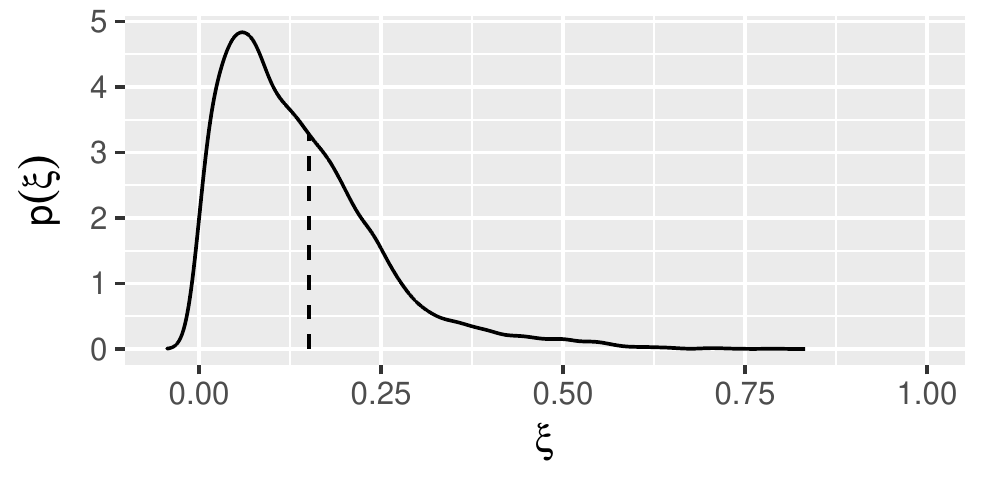}}\\
\subfloat[R7]{\includegraphics[scale=0.6]{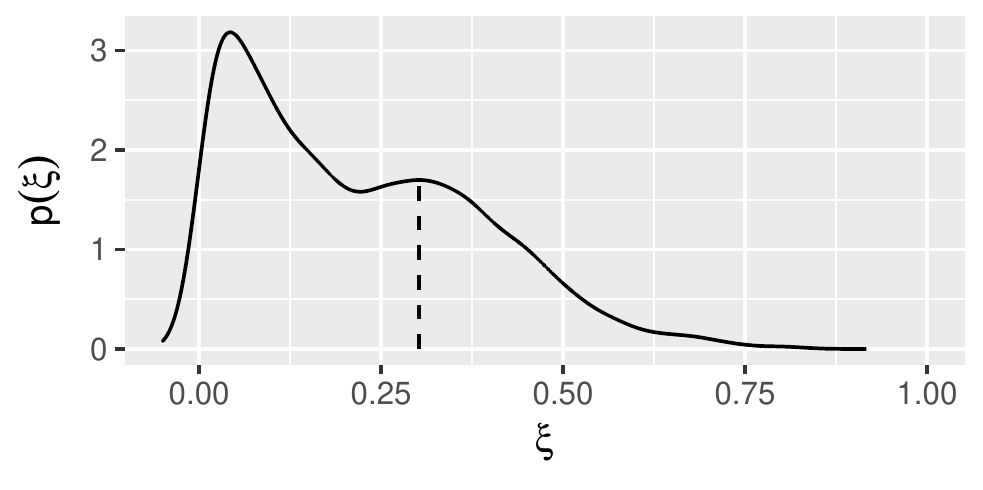}} & \\
\end{tabular}
\caption{Empirical density plots of $\xi$, the proportion of epialleles at a locus that are different from normal tissue. The distribution of $\xi$ will depend on the tumour purity since samples that are contaminated with less normal tissue will exhibit a greater deviance from the matched normal tissue epialleles. On this basis, the rightmost maxima (marked with a dashed vertical line) of the empirical densities are interpreted as a proxy for sample purity.}
\label{fig:purity}
\end{figure}

\newpage
\section{Supplementary figures}
\begin{figure}[h!]
\centering
\includegraphics[scale=0.6]{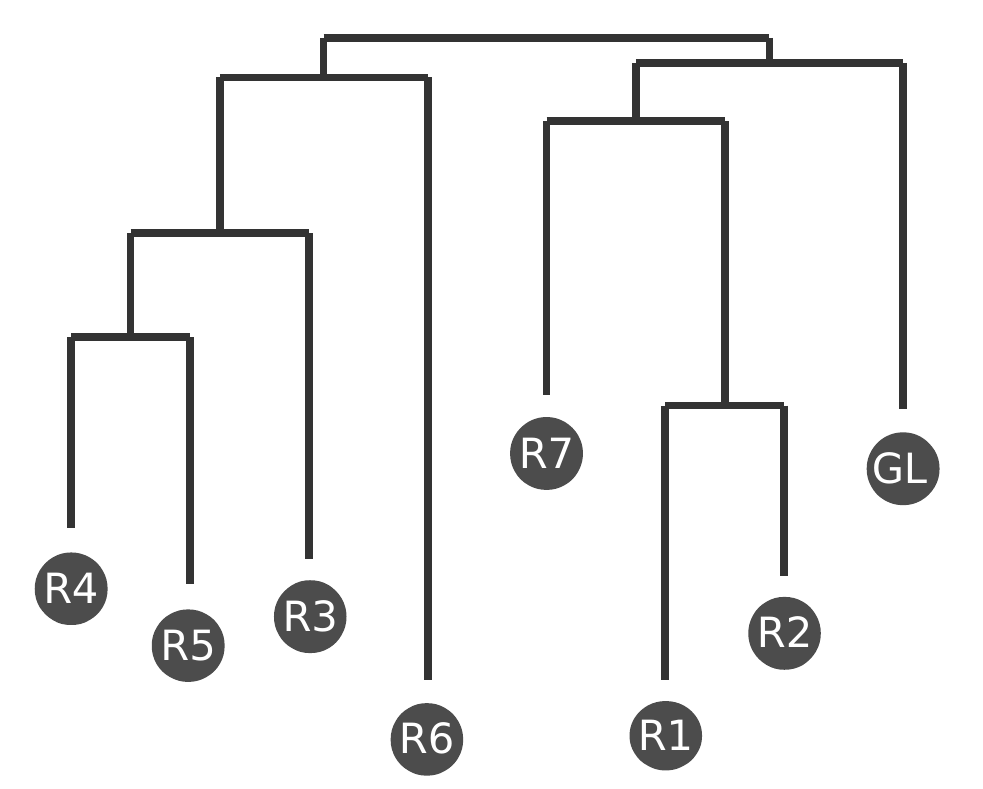}
\caption{Phylogenetic tree generated from exome sequencing data from the same tumour that is studied in the main text. The exome data were generated and analysed independently as part of the Jamal-Hanjani et al., 2017 study. GL denotes germline.}
\label{fig:example}
\end{figure}

\section{Experimental datasets}

The RRBS data are available at the European Nucleotide Archive under accession numbers ERS1546024, ERS1546025 and ERS1546026.

\newpage
\section{TRACERx consortium members}
%
%
%

The TRACERx study (Clinicaltrials.gov no: NCT01888601) is sponsored by University College London (UCL/12/0279) and has been approved by an independent Research Ethics Committee (13/LO/1546). TRACER is funded by Cancer Research UK (grant number C11496/A17786) and coordinated through the Cancer Research UK \& UCL Cancer Trials Centre.

\subsection*{Consortium members}

Charles Swanton$^{1,2,5}$, Mariam Jamal-Hanjani$^{1}$, Selvaraju Veeriah$^{1}$, Seema Shafi$^{1}$, Justyna Czyzewska-Khan$^{1}$, Diana Johnson$^{1}$, Joanne Laycock$^{1}$, Leticia Bosshard-Carter$^{1}$, Gerald Goh$^{1}$, Rachel Rosenthal$^{1}$
% ...

\subsection*{Affiliations}
\setlength{\parindent}{0pt}
1. Cancer Research UK Lung Cancer Centre of Excellence, University College London Cancer Institute, United Kingdom

2. The Francis Crick Institute, United Kingdom

3. Lungs for Living, UCL Respiratory, University College London, United Kingdom

4. The Royal Marsden Hospital, United Kingdom

5. University College London Hospitals NHS Foundation Trust, United Kingdom

6. University Hospital of South Manchester, United Kingdom

7. Manchester Cancer Research Centre Biobank, United Kingdom

8. Christie NHS Foundation Trust, Manchester, United Kingdom

9. Cancer Research UK Manchester Institute, United Kingdom

10. Heart of England NHS Foundation Trust, Birmingham, United Kingdom

11. Cancer Studies and Molecular Medicine, University of Leicester, United kingdom

12. Leicester University Hospitals, United Kingdom

13. National Institute for Health Research Leicester Respiratory Biomedical, Research Unit, United Kingdom

14. North Middlesex Hospital, United Kingdom

15. Royal Free Hospital, United Kingdom

16. Barnet Hospital, United Kingdom

17. Aberdeen Royal Infirmary, United Kingdom

18. Velindre Cancer Centre, Cardiff, Wales, United Kingdom

19. Cardiff \& Vale University Health Board, Cardiff, Wales, United Kingdom

20. University Hospital Of Wales Heath Park, Cardiff, Wales, United Kingdom

21. Department of Pathology, University Hospital of Wales and Cardiff University, Heath Park, Cardiff, Wales, United Kingdom

22. The Whittington Hospital NHS Trust, United Kingdom

23. Independent Cancer Patients Voice, United Kingdom

24. Cancer Research UK \& UCL Cancer Trials Centre, United Kingdom

25. University Hospital Southampton NHS Foundation Trust, United Kingdom

26. Royal Brompton and Harefield NHS Foundation Trust, United Kingdom

27. National Heart and Lung Institute, Imperial College, United Kingdom

28. Barts Health NHS Trust, United Kingdom

29. Ashford and St. Peter's Hospitals NHS Foundation Trust, United Kingdom

30. The Princess Alexandra Hospital NHS Trust, United Kingdom

31. Sheffield Teaching Hospitals NHS Foundation Trust, United Kingdom

32. Liverpool Heart and Chest Hospital NHS Foundation Trust, United Kingdom

33. Natera Inc., 201 Industrial Road, Suite 410, San Carlos, CA 94070

34. Dana-Farber Cancer Institute, Brigham \& Women's Hospital, Harvard Medical School, 450 Brookline Ave, JF518, Boston, MA 02115-5450, USA

35. Wellcome Trust Sanger Institute, Hinxton, CB10 1SA, United Kingdom

36. Bioinformatics Group, Department of Computer Science, University College London

37. University of Lausanne

38. Swiss Institute of Bioinformatics

\end{document}